# F.A.S.T.

## Floor field- and Agent-based Simulation Tool

Tobias Kretz and Michael Schreckenberg
{kretz, schreckenberg}@traffic.uni-duisburg.de
Universität Duisburg-Essen
Lotharstr. 1, 47048 Duisburg

*Abstract*

In this work a model of pedestrian motion is presented. As application its parameters are fitted to one run in a primary school evacuation exercise. Simulations with these parameters are compared to further runs during the same exercise.

## Introduction

Understanding the dynamics of crowds in different situations has gained increasing interest over the last decades. Whoever carefully watches the international media will note roughly one incident in junction with crowds with disastrous or almost-disastrous results per month world-wide [1].

On this background it is not surprising that almost as long as large crowds and people in large buildings have been recognized to possibly hold some special problems, calculation frameworks have been set up to estimate evacuation times and other crucial values of pedestrian motion.

The complexity of the models increased with the available calculational power. The road went from hand-calculation [2] and hydrodynamic models over network models to models with an individual representation (agents, "microscopic simulation") of real persons and an ever more exact inclusion of the environment, sometimes with a discrete [3],[4], sometimes with a continuous [5] representation of space and time.

## A Model of Pedestrian Motion

In this section a model of pedestrian motion, that is implemented into the F.A.S.T. (<u>F</u>loor field- and <u>A</u>gentbased <u>S</u>imulation <u>T</u>ool), is presented that is microscopic and that to a large extend makes use of so called floor fields (compare [6]) to determine the motion of the agents. The model is discrete in space and time. The agents move on a grid of cells which represent squares of 40•40 cm². The time advances in rounds. Each round is interpreted as one second. Each cell can at maximum be occupied by one single agent.

Floor fields fulfil two tasks:
1) Constant floor fields, in a simple way allow to save calculation time, as important values like the distance of a cell to an exit are saved in them.
2) Floor fields that change with time, can be used to transform long-ranged interactions into short-ranged ones. This is a more sophisticated element to save calculation time.

The F.A.S.T. model contains three floor fields:

1) The "Static floor field" (compare [3], [7]) for each cell contains the information of the distance to the exit. In fact there is one static floor field for each exit. (Exit-cells that are connected by a common edge are grouped to exits.) The static floor field can be understood as some kind of potential in which the agents "fall" towards the exit.
2) The "Dynamic floor field" (compare [3], [8]) is a vector field. An agent who has moved from cell *(a, b)* to cell *(x, y)* changes the dynamic floor field *($D_x$, $D_y$)* at *(a, b)* by *(x-a, y-b)* after all agents have moved. The dynamic floor field does not change on intermediate cells, which the agents cross on their way from their source to their destination cell. Right after that all values of both components of *D* decay with probability *δ* and diffuse with probability *α* to one of the (von Neumann) neighbouring cells. Since the vector components can be negative, decay means a reduction of the absolute value. Diffusion is only possible from x- to x- and from y- to y-component. Diffusion from a negative valued component means lowering the component value at the target cell whether it is positive or negative and vice versa for positive values.
3) The distance of a cell to the next wall is also saved in a floor field if this distance is smaller than a certain threshold. This is a simple construction to avoid calculating the distance every time it is needed.

There are other influences on the motion that are not governed by floor fields but rather by properties of the agents:
1) There is an inertia that makes the agents avoid sharp turns at high velocities. This inertia is not the normal inertia of Newtonian physics. Due to the special construction of the human movement apparatus a person can relatively easily accelerate or decelerate into the normal direction of motion, while deviating from that direction strongly on short time-scales is more difficult, especially at high velocities.
2) If possible it is typically avoided to move too close to other people, and so an agent also can have a repulsive effect on other agents – if enough cells are available that are not close to agents.
3) Some kind of "friction" (compare [10]) is implemented that reduces the effectiveness with which agents, that compete for the same cell during one round, reach that cell.
4) If the scenario includes more than one exit, the agents choose one of them at the beginning of each round. Here the decision of the last round plays an important role, as a once taken decision typically is not revised each second.

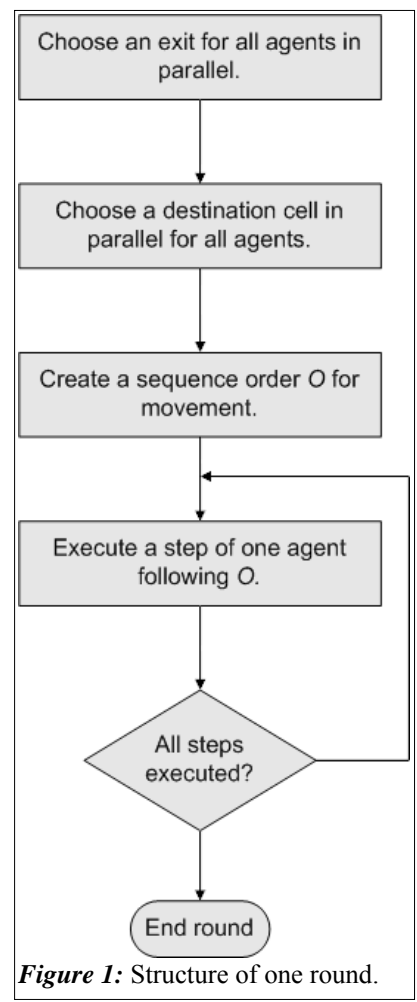

*Figure 1:* Structure of one round.

Except for the last two influences, the strength of all influences is determined by coupling constants: An agent "couples" to the static floor field, to the dynamic floor field, to the wall-field, to his own inertia and to the presence of other agents. All of these coupling constants can be interpreted in some way. The coupling to the static floor field can be interpreted as the knowledge the agent has of his environment, the coupling to his own inertia as ratio strength to body mass or more generally his fitness.

Figure 1 shows how these influences are merged into a three-phase process in each round. At first all agents choose the exit they want to approach during that particular round. According to the influences described above and to Equation 1, in the second phase all agents in parallel choose a destination cell out of all cells they can reach theoretically during that round [9]. Walls and cells that are occupied by other agents are excluded. The set of reachable cells is determined by the personal maximal speed of an agent. In the last phase all agents move and try to reach their destination cell. This they do not necessarily accomplish if other agents intercept the path to the destination cell.

$$p_{xy} = N p_{xy}^S p_{xy}^D p_{xy}^I p_{xy}^W p_{xy}^P$$

*Equation 1:* Probability for cell *(x,y)* to be chosen as destination cell. The details of the influences from the static ($p^S$) and dynamic ($p^D$) floor field, as well as from inertia ($p^I$), the walls ($p^W$), and other agents ($p^P$) are explained in Appendix A.

Up to now F.A.S.T. has been validated at a number of simple scenarios [11] and results of an evacuation exercise at a primary school have been compared to the results of a F.A.S.T Simulation of this exercise. Concerning the simple scenarios F.A.S.T. brought up results that are comparable to a still widely used hand-calculation method [2], implying that the evacuation times predicted by F.A.S.T. are typically but not always more conservative than those predicted by some commercially available software packages [12], [13], [14], [15]. Additionally the F.A.S.T. model has been used as an example to study oscillations at narrow bottlenecks [16]. Out of these in the following a detailed report on the evacuation exercise at a primary school shall be given.

## Evacuation Exercise in a Primary School

The evacuation exercise that was reported about in [17] has been repeated. However this time there were fewer pupils. The children were highly motivated, which was partly due to the presence of a camera team reporting for a children's news show on a German children's TV station. The exercise was repeated twice. The first time the music class did not become aware of the alarm as the bell in their class-room was broken and they were singing too loud to hear the bell from the floor. The school consists of two buildings: The main building and a newer second building. The music class was on the second floor in the main building. The main building consists of three (first, second and third floor), the second building of two floors (first and second floor). See Figure 2.

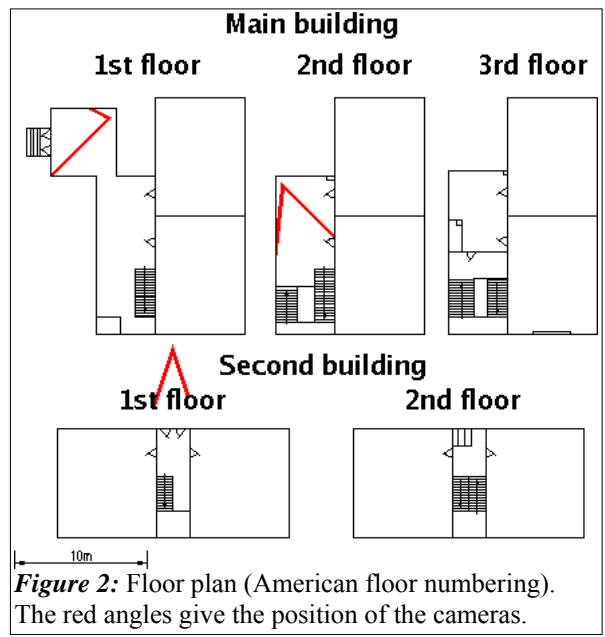

*Figure 2:* Floor plan (American floor numbering). The red angles give the position of the cameras.

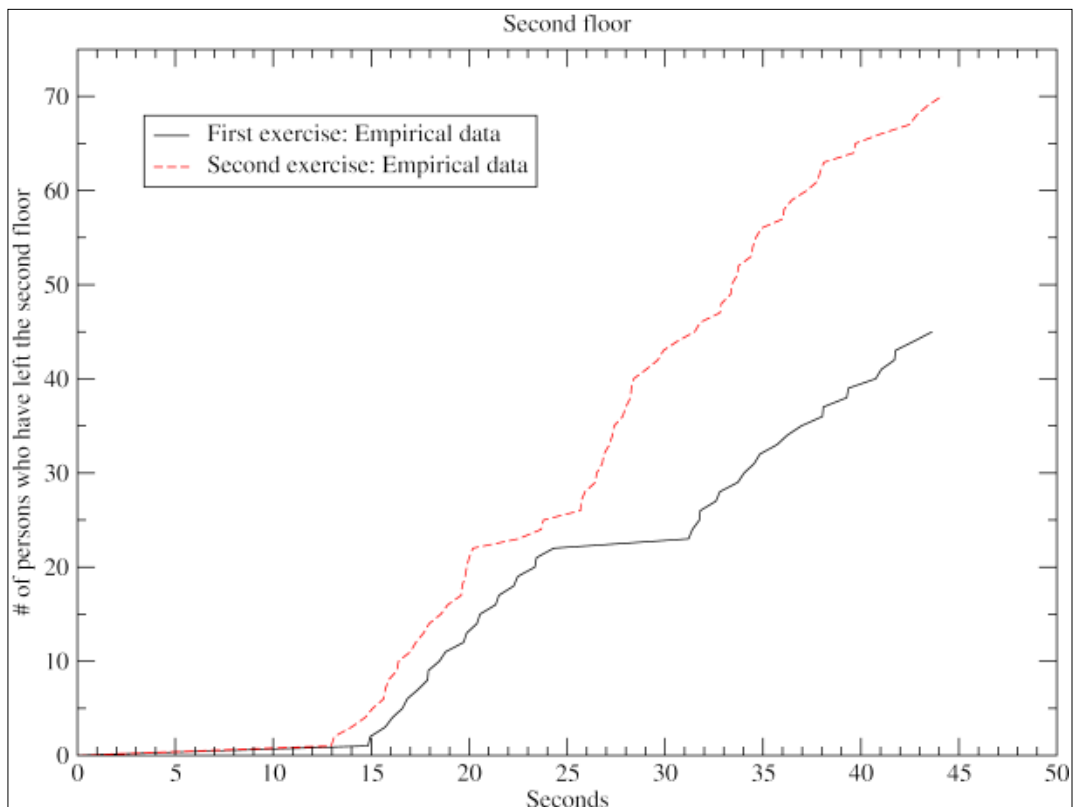

*Figure 3:* Evacuation graphs of both runs in the second floor of the main building. The pupils were counted, when they left the second floor and moved down the first step of the stairway between second and first floor. In the first run (without the music class) the pupils of the other than the music class on the second floor had left the second floor some time before the pupils of the third floor arrived. In the second run the pupils from the third floor arrived before all pupils of the second floor had left the second floor, but there was some dawdling of two pupils without apparent reason, leading again to an - in this case smaller - plateau in the evacuation graph.

In addition to the three cameras the time for the last person to leave the third floor was measured. A person was counted as having exited the main building as he reached the last of the stairs outside the main building.

*Results*

| *Table 1*: Results: Time (in seconds) after alarm for... | First Exercise | Second Exercise |
|---|---:|---:|
| ...the last person to leave the main building. | 65.4 | 69.9 |
| ...the first person to leave the main building. | 28.5 | 12.3 |
| ...the last person to leave the 2nd floor of the main building. | 43.2 | 44.9 |
| ...the first person to leave the 2nd floor of the main building. | 15.3 | 13.2 |
| ...the last person to leave the 3rd floor of the main building. | 25.0 | 24.0 |
| ...the last person to leave the second building. | 60.5 | 56.5 |
| ...the first person to leave the second building. | 16.2 | 5.2 |

The results (Table 1) of the two exercises in the main building can hardly be compared since the music class only took part in the second exercise. The data of the second building however suggest that in the second run there either was a learning effect or that the pupils - at least some of them - were more aware of an alarm to come, instead of having normal lessons interrupted by an alarm, since they reacted more quickly. See Figure 3 and Figure 4.

*Comparison to Simulation Results*

After the exercise was finished and the empirical data was evaluated, simulations were done with the aim to reproduce the empirical data of the first exercise as well as possible. This resulted in the evacuation graphs of Figure 5. Due to technical reasons an empirical evacuation graph at the main exit could not be evaluated, but the total time - averaged over 1000 simulation runs - of the evacuation (until all pupils had completely left the main building) was *62.2* seconds at a standard deviation of *1.3* seconds. The smallest evacuation time that appeared during those 1000 simulation runs was *58* seconds, the largest *69* seconds. For the evacuation of the second floor those numbers were: *44.6 ± 1.4* seconds with all evacuation times between *42* and *52* seconds.

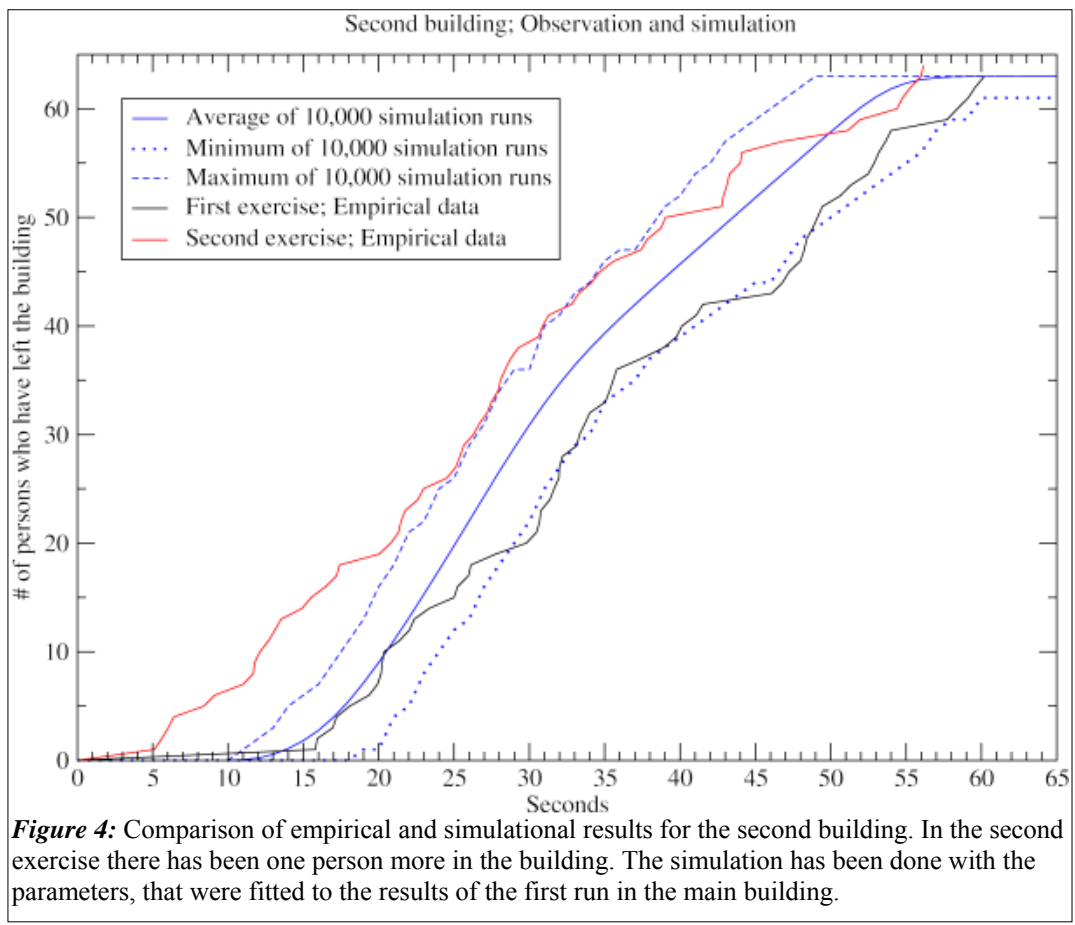

*Figure 4:* Comparison of empirical and simulational results for the second building. In the second exercise there has been one person more in the building. The simulation has been done with the parameters, that were fitted to the results of the first run in the main building.

The parameters that led to the results of Figure 5 were as follows (see equations 2 – 7 in appendix A for an explanation of the parameters): $k_S=3.0$, $k_D=2.0$, $k_I=2.0$, $k_W=0$, trace strength: 6, $\alpha=0.8$, $\delta=0.5$, $\mu=k_P=0$. For the reaction times of the teachers and the pupils on the third floor (fourth grade, oldest pupils of the school) the following distribution of reaction times was used: $t_r^{min}=18$ seconds, $t_r^{av}=19$ seconds, $t_r^{max}=20$ seconds, $t_r^{std}=1$ second. The maximum speed was set to $v_{max}=5$

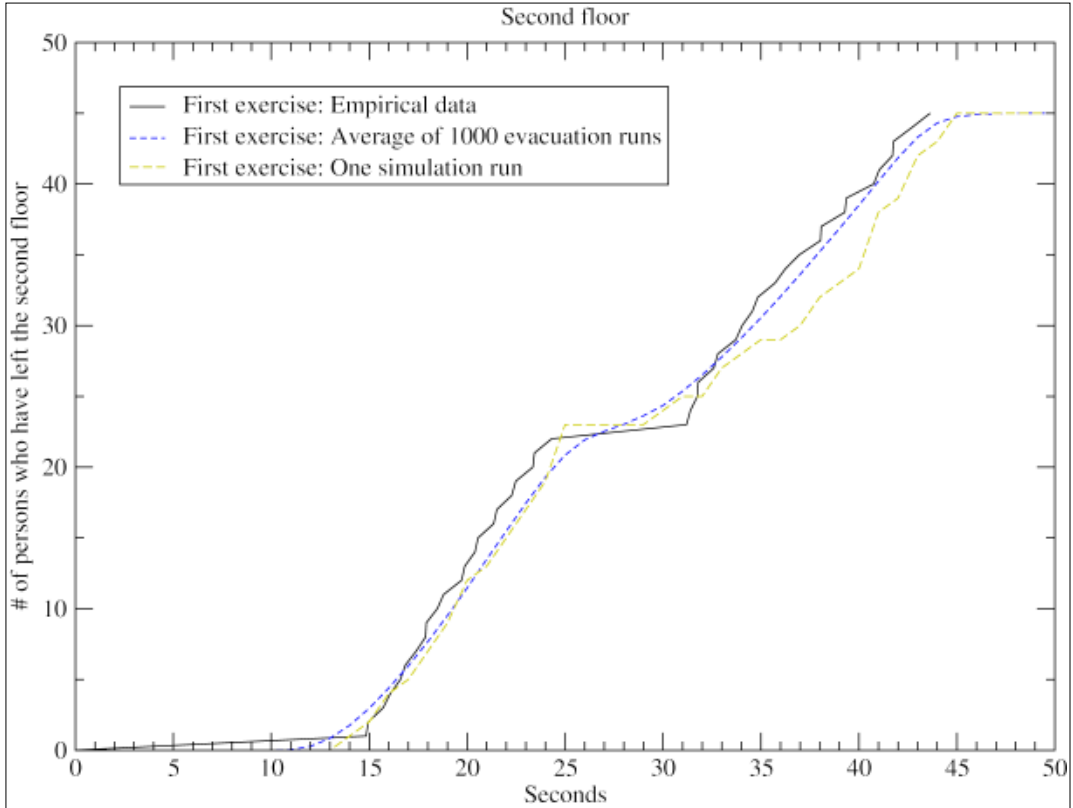

*Figure 5:* Comparison of empirical and simulational results. The simulations were done after the exercise, so this is not a prediction but a calibration of the simulation.

(cells per round) for all of them, while for the other (younger pupils) the reaction time was set on smaller values for some of them $t_r^{min}=10$ seconds, $t_r^{av}=15$ seconds, $t_r^{max}=20$ seconds, $t_r^{std}=5$ seconds and the speed varied: $v_{max}^{min}=4$, $v_{max}^{av}=6$, $v_{max}^{max}=8$, $v_{max}^{std}=1$. This corresponds to the following observations: Some of the younger pupils were highly motivated, speeds up to 3 meter per second were observed. The older students of the third floor stayed slightly closer together and appeared to be slightly less (but still highly) motivated and/or more disciplined. It might surprise that all pupils seem to have such a strong inertia, but $k_I$ always has to be set and seen in relation to $k_S$ and it was indeed the case that the turnaround on the stairway slowed the pupils significantly down. Note: Even small variations in some parameters as the maximum speed, the reaction times, $\alpha$, $\delta$, the trace strength, $k_D$, $k_I$ and to some extent $k_S$ lead to a much smaller agreement between observation and simulation. It was especially difficult to find parameters that reproduce the plateau in the evacuation graph.

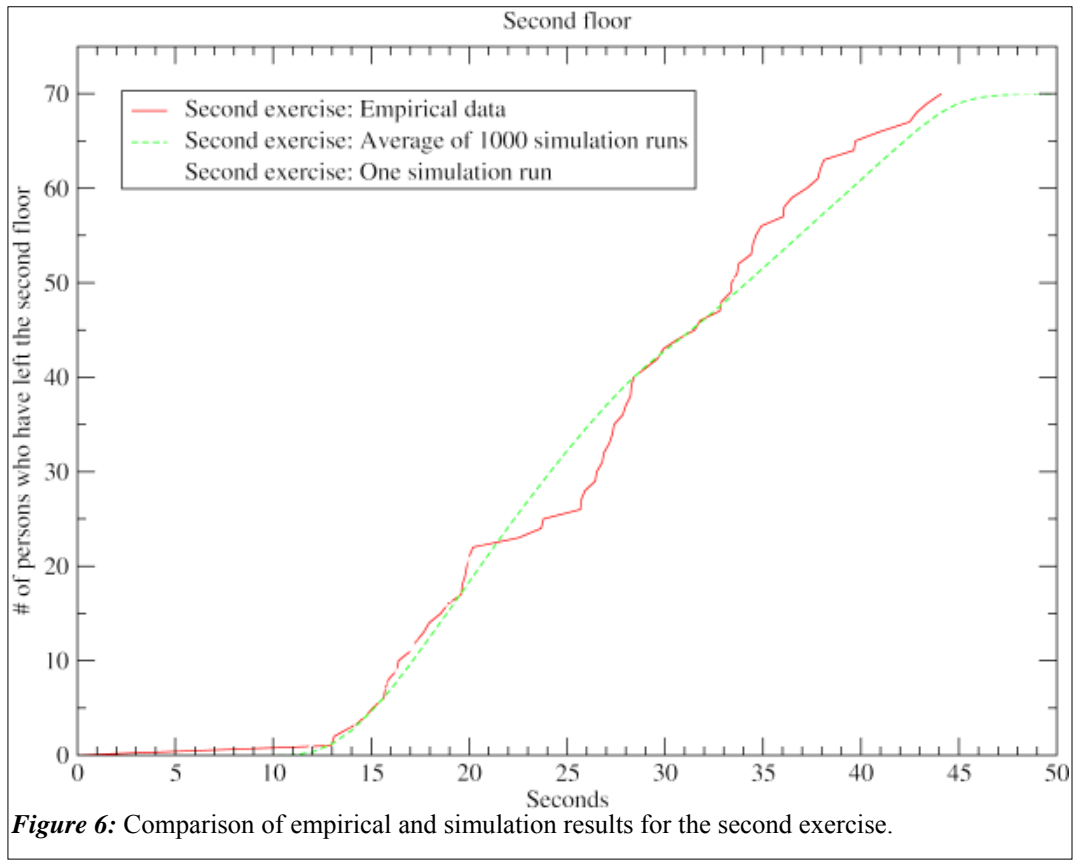

*Figure 6:* Comparison of empirical and simulation results for the second exercise.

Now these parameters have been used in simulations that include the music class (second run). A comparison of observation and simulation is shown in Figure 6. The total time - averaged over 1000 simulation runs - of the evacuation is *67.7* seconds at a standard deviation of *1.7* seconds. The smallest evacuation time that appeared during those 1000 simulation runs was *63* seconds, the largest *75* seconds. For the evacuation of the second floor those numbers were: *46.0 ± 1.7* seconds with all evacuation times between *42* and *56* seconds.

While the parameters have been calibrated at the data of the evacuation of the second floor and the first exercise, the results of the simulation for the evacuation of the whole building at the second exercise (*67.7 ± 1.7* seconds, minimum *63*, maximum *75* seconds) are also in good agreement with the corresponding empirical data (*69.9* seconds). The fact that no set of parameters could be found that fully reproduces the high outflow from the second floor is probably due to the smaller size of the children compared to adults for which normally data is taken in experiments and observations.

The same parameters applied to the second building led to an average simulated evacuation time of *56.0 ± 2.2* seconds. (See Figure 4.) Compared to the first exercise the students in the second building performed better throughout the whole second exercise. It is not clear is this was because they were more aware of another alarm to come or if they actually performed better during egress, due to a practicing effect. The average of simulated evacuation times yields results that are almost identical to the results of the first exercise at the beginning of the process and results that fit very well to the data of the second exercise at the end of the process.

## Summary

In this work a model of pedestrian motion was presented. As an example of parameter gauging results of the model were compared to an evacuation exercise in a primary school. Within the variations between the two exercises the results of the simulation are in good agreement with the empirical data. However the agreement is better for identical floor plan and different population number than for identical population number and different floor plan.

## Acknowledgments

This work was financed by the Bundesministerium für Bildung und Forschung (BMBF) within the PeSOS project. We thank Ms. Dommers, headmaster of the primary school ``Am Knappert´´ in Duisburg-Rahm, and her colleagues for their cooperation, furthermore Anna Grünebohm, Frank Königstein, Florian Mazur, and Mareike Quessel for their support during the evacuation exercise, respectively the evaluation process.

# Appendix A: Formulae

$$p_E^A = N \frac{1 + \delta_{AE} k_E(A)}{S(A,E)^2}$$

*Equation 2:* Probability that agent A chooses exit E. $\delta_{AE} = 1$ for the exit that was chosen by agent A at the last round, $k_E$ is a coupling constant, that has to be set to fit the simulation to the circumstances of an evacuation.

$$p_{xy}^S = e^{-k_S S_{xy}^e}$$

*Equation 3:* Partial probability from the static floor field S, that cell (x, y) is chosen as destination cell. $S^e_{xy}$ contains the information on the distance of cell (x, y) to exit e, $k_S$ is a coupling constant, that has to be set to fit the simulation to the circumstances of an evacuation.

$$p_{xy}^D = e^{k_D(D_x(x,y)(x-a) + D_y(x,y)(y-b))}$$

*Equation 4:* Partial probability from the static floor field D, that cell (x, y) is chosen as destination cell. (a,b) is the current position of agent A and (x, y) the position of the cell in focus. $k_D$ is a coupling constant, that has to be set to fit the simulation to the circumstances of

$$p^I(\Delta x_{t+1}, \Delta y_{t+1}) = \exp\left(-k_I(v_{next} + v_{last})\sqrt{\frac{1}{2}\left(1 - \frac{\begin{pmatrix}\Delta x_{t+1}\\ \Delta y_{t+1}\end{pmatrix}\begin{pmatrix}\Delta x_t\\ \Delta y_t\end{pmatrix}}{\left|\begin{pmatrix}\Delta x_{t+1}\\ \Delta y_{t+1}\end{pmatrix}\right|\left|\begin{pmatrix}\Delta x_t\\ \Delta y_t\end{pmatrix}\right|}\right)}\right)$$

*Equation 5:* Partial probability from the inertia, that cell (x, y) is chosen as destination cell. $(\Delta x_t, \Delta y_t)$ is the velocity vector of the last round. $(\Delta x_{t+1}, \Delta y_{t+1})$ the possible velocity vector of this round (the vector about which still has to be decided,. $v_{last}$ and $v_{next}$ the corresponding absolute values (rounded to integers). $k_I$ is a coupling constant, that has to be set to fit the simulation to the circumstances of an evacuation. A derivation of this equation can be found in

$$p_{xy}^W = e^{(-k_W W_{xy})}$$

*Equation 6:* Partial probability from possible nearby walls, that cell (x, y) is chosen as destination cell. $W_{xy}$ is $W_0$ minus the distance to the wall closest to agent A. If $W_{xy}$ is larger than the cut-off parameter $W_0$, $W_{xy}$ is set to 0. $k_W$ is a coupling constant, that has to be set to fit the simulation to the circumstances of an evacuation.

$$p_{xy}^P = e^{-k_P N_P(x,y)}$$

*Equation 7:* Partial probability from possible nearby agents, that cell (x, y) is chosen as destination cell. $N_P$ is the number of cells within the Moore neighbourhood of (x, y) occupied by other agents. $k_P$ is a coupling constant, that has to be set to fit the simulation to the